\documentclass[aps,prd,reprint,preprintnumbers,showpacs,floatfix,nofootinbib,superscript address,longbibliography]{revtex4-2}
\usepackage[utf8]{inputenc}
\usepackage{amssymb}
\usepackage{hhline}
\usepackage{amsmath}
\usepackage{mathtools}
\usepackage[dvipsnames]{xcolor}
\usepackage{multirow,tabularx}
\usepackage{siunitx}
\usepackage{graphicx}
\usepackage{natbib}

\usepackage{etoolbox}
\usepackage{tensor}

\newrobustcmd{\PSALTer}{\textit{PSALTer}\xspace}
\newrobustcmd{\MPl}{%
	{M_{\mathrm{P}}} 
}
\newrobustcmd{\Danger}[1]{%
	{\color{red}{#1}}	
}
\newrobustcmd{\Coupling}[1]{%
	{a_{#1}}	
}
\newrobustcmd{\CouplingB}[1]{%
	{b_{#1}}	
}
\newrobustcmd{\MAGg}[1]{%
	\tensor{g}{#1}
}
\newrobustcmd{\MAGd}[1]{%
	\tensor*{\delta}{#1}
}
\newrobustcmd{\MAGl}[1]{%
	\tensor{\xi}{#1}
}
\newrobustcmd{\MAGP}[1]{%
  \tensor{P}{#1}
}
\newrobustcmd{\MAGC}[1]{%
  \tensor{C}{#1}
}
\newrobustcmd{\ShiftA}[1]{%
  \tensor{A}{#1}
}
\newrobustcmd{\ShiftB}[1]{%
  \tensor{B}{#1}
}
\newrobustcmd{\ShiftC}[1]{%
  \tensor{C}{#1}
}
\newrobustcmd{\MAGA}[1]{%
  \tensor{\Gamma}{#1}
}
\newrobustcmd{\Ja}[1]{%
	\tensor{\smash{\overset{\scalebox{0.6}{$\mathrm{(M)}$}}{\mathcal{J}}}}{#1}
}
\newrobustcmd{\Jv}[1]{%
	\tensor{\smash{\overset{\scalebox{0.6}{$\mathrm{(N)}$}}{\mathcal{J}}}}{#1}
}
\newrobustcmd{\MAGF}[1]{%
	\tensor{R}{#1}
}
\newrobustcmd{\MAGFt}[1]{%
	\tensor{\tilde{R}}{#1}
}
\newrobustcmd{\MAGFh}[1]{%
	\tensor{\hat{R}}{#1}
}
\newrobustcmd{\MAGFa}[1]{%
	\tensor{\check{R}}{#1}
}
\newrobustcmd{\MAGFb}[1]{%
	\tensor{R}{#1}
}
\newrobustcmd{\MAGT}[1]{%
	\tensor{T}{#1}
}
\newrobustcmd{\MAGTt}[1]{%
	\tensor{\hat{T}}{#1}
}
\newrobustcmd{\MAGt}[1]{%
	\tensor{t}{#1}
}
\newrobustcmd{\MAGQ}[1]{%
	\tensor{Q}{#1}
}
\newrobustcmd{\MAGQt}[1]{%
	\tensor{\hat{Q}}{#1}
}
\newrobustcmd{\MAGq}[1]{%
	\tensor{q}{#1}
}
\newrobustcmd{\MAGV}[1]{%
	\tensor{Z}{#1}
}
\newrobustcmd{\MAGVt}[1]{%
	\tensor{\tilde{Z}}{#1}
}

\newrobustcmd{\xMAGA}[1]{%
  \tensor{A}{#1}
}
\newrobustcmd{\xMAGF}[1]{%
	\tensor{\mathcal{F}}{#1}
}
\newrobustcmd{\xMAGFt}[1]{%
	\tensor{\tilde{\mathcal{F}}}{#1}
}
\newrobustcmd{\xMAGFa}[1]{%
	\tensor{\mathcal{F}}{^{(14)}#1}
}
\newrobustcmd{\xMAGFb}[1]{%
	\tensor{\mathcal{F}}{^{(13)}#1}
}
\newrobustcmd{\xMAGT}[1]{%
	\tensor{\mathcal{T}}{#1}
}
\newrobustcmd{\xMAGTt}[1]{%
	\tensor{\tilde{\mathcal{T}}}{#1}
}
\newrobustcmd{\xMAGQ}[1]{%
	\tensor{\mathcal{Q}}{#1}
}
\newrobustcmd{\xMAGQt}[1]{%
	\tensor{\tilde{\mathcal{Q}}}{#1}
}
\newrobustcmd{\xMAGTh}[1]{%
	\tensor{\hat{\mathcal{T}}}{#1}
}
\newrobustcmd{\xMAGV}[1]{%
	\tensor{\mathcal{V}}{#1}
}

\newrobustcmd{\g}[1]{%
	\tensor{g}{#1}%
}
\newrobustcmd{\rcCon}[1]{%
	\tensor*{\Gamma}{#1}%
}
\newrobustcmd{\rCon}[1]{%
	\tensor*{\mathring{\Gamma}}{#1}%
}
\newrobustcmd{\Con}[1]{%
	\tensor{\mathring{\Gamma}}{#1}%
}
\newrobustcmd{\B}[1]{%
	\tensor{B}{#1}%
}
\newrobustcmd{\PD}[1]{%
	\tensor{\partial}{#1}%
}
\newrobustcmd{\rD}[1]{%
	\tensor{\mathring{\nabla}}{#1}%
}
\newrobustcmd{\rcD}[1]{%
	\tensor{\nabla}{#1}%
}
\newrobustcmd{\rR}[1]{%
	\tensor{\mathring{R}}{#1}%
}
\newrobustcmd{\rcR}[1]{%
	\tensor{R}{#1}%
}

\usepackage{xspace}
\usepackage{xstring}
\usepackage{titlesec}
\usepackage{parskip}
\usepackage{stix}
\allowdisplaybreaks
\newcommand*{\ie}{i.e., }
\newcommand*{\eg}{e.g., }

\newcommand*{\eqs}{eqs.\@\xspace}

\newrobustcmd{\pea}[1]{%
	\emph{#1}\textbf{\ \ \ ---}
}
\titleformat{\paragraph}[runin]{\normalfont\normalsize\bfseries}{\emph\theparagraph}{1em}{\pea}

\usepackage{hyperref}
\hypersetup{%
     colorlinks = true,%
     linkcolor = Blue,%
     citecolor = Blue,%
     filecolor = Blue,%
     urlcolor = Blue% 
     }%
\usepackage[capitalize,nameinlink]{cleveref}

\begin{document}

\title{Consistent particle physics in metric-affine gravity from extended projective symmetry}

\author{Will Barker}
\email{wb263@cam.ac.uk}
\affiliation{Astrophysics Group, Cavendish Laboratory, JJ Thomson Avenue, Cambridge CB3 0HE, UK}
\affiliation{Kavli Institute for Cosmology, Madingley Road, Cambridge CB3 0HA, UK}
\author{Sebastian Zell}
\email{sebastian.zell@uclouvain.be}
\affiliation{Centre for Cosmology, Particle Physics and Phenomenology -- CP3, Universit\'e catholique de Louvain, B-1348 Louvain-la-Neuve, Belgium}

%\date{}

\begin{abstract}
	It is well-known that the Einstein-Hilbert action exhibits a projective invariance in metric-affine gravity, generated by a single vector (just like diffeomorphisms). However, this symmetry offers no protection against formulating inconsistent models, \eg with ghost and strong coupling problems. In this letter, we observe that non-minimal kinetic terms of Dirac spinors point to a new \emph{extended projective} (EP) symmetry generated by a pair of vectors. We prove that the most general EP-invariant theory (at most quadratic in field strengths) is naturally free from all pathologies. Its spectrum only features the massless graviton and a single additional scalar field arising from the square of the Holst curvature. The scalar potential is suitable for inflation and our model moreover contains effective 4-Fermi interactions capable of producing fermionic dark matter. Finally, we point out an alternative double-vector symmetry that similarly leads to a healthy theory with a propagating vector field.
\end{abstract}

\maketitle

\paragraph*{Fundamental forces from symmetries} Among the greatest achievements of fundamental science is a description of our world in terms of as few as four basic forces. Three of them -- the strong, weak, and electromagnetic interactions -- are at the heart of the Standard Model (SM) of particle physics and are crucial for understanding phenomena on microscopic scales. The fourth force, gravity, is carried in by the theory of General Relativity (GR) and shapes the macroscopic structure of our Universe. All four forces are fundamentally defined in terms of symmetries. On the particle physics side, the essence of the SM is captured by the internal gauge groups~$\mathrm{SU}(3)$,~$\mathrm{SU}(2)$, and~$\mathrm{U}(1)$ corresponding to the strong, weak, and electromagnetic interactions, respectively. In gravity, invariance under external spacetime diffeomorphisms allows local observers to choose their own coordinate systems.
	
\paragraph*{Constraints on possible theories} One can understand the key role of symmetries as constraints on possible theories. Since only invariant terms (operators) are admissible in the action, the theory space is severely reduced. Famously, the symmetries of the SM forbid the existence of explicit mass terms. This led to the seminal theoretical prediction of the Brout-Englert-Higgs mechanism~\cite{Englert:1964et,Higgs:1964pj}, which, decades later, was confirmed with the discovery of the Higgs boson~\cite{CMS:2012qbp,ATLAS:2012yve}. Compared to the SM, the situation in GR is more ambiguous. This is because diffeomorphism invariance does not uniquely determine the fundamental geometry of spacetime: the presence of the three geometric properties curvature, torsion and non-metricity define various equivalent formulations of GR~\cite{Einstein:1915,Weyl:1918, Palatini:1919,Weyl:1922,Cartan:1922, Eddington:1923,  Cartan:1923,Eddington:1923, Cartan:1924, Cartan:1925,Einstein:1925, Einstein:1928, Einstein:19282,Schroedinger:1950,Moller:1961, Pellegrini:1963, Hayashi:1967se,Cho:1975dh,Hehl:1976kt, Hehl:1976kv, Hehl:1976my, Hehl:1977fj,Kijowski:1978,Hayashi:1979qx, Nester:1998mp, BeltranJimenez:2019odq}; see~\cite{Heisenberg:2018vsk,BeltranJimenez:2019esp,Rigouzzo:2022yan,Heisenberg:2023lru} for overviews. Might this ambiguity indicate that more symmetries of gravity are yet to be discovered?

\paragraph*{Metric-affine gravity} Without ad hoc assumptions about the geometry of spacetime, the field content of GR comprises a metric~$\tensor{g}{_{\mu\nu}}$ and an independent affine connection~$\MAGA{^\mu_\nu_\rho}$. The latter can be field-reparameterised into torsion and non-metricity tensors (see conventions in~\cref{app:conventions})
\begin{equation}
	\MAGT{^\alpha_\mu_\nu}\equiv 2\MAGA{^\alpha_{[\mu}_{\nu]}} \;,\quad
	\MAGQ{_{\lambda\mu\nu}}\equiv \rcD{_\lambda}\MAGg{_{\mu\nu}}\;,\label{tqdef}
\end{equation}
where~$\rcD{_\lambda}\MAGg{_{\mu\nu}}\equiv\tensor{\partial}{_\lambda}\MAGg{_{\mu\nu}}-2\MAGA{^\alpha_{\lambda(\mu}}\MAGg{_{\nu)\alpha}}$. The curvature is
\begin{equation}
	\MAGF{^\rho_\sigma_{\mu\nu}}\equiv 2\left(\tensor{\partial}{_{[\mu|}}\MAGA{^\rho_{|\nu]}_\sigma}+\MAGA{^\rho_{[\mu|}_\alpha}\MAGA{^\alpha_{|\nu]}_\sigma}\right),\label{FDef}
\end{equation}
 and~\cref{tqdef,FDef} determine the effects of parallel transport. Allowing for~$\MAGF{^\rho_\sigma_{\mu\nu}}$,~$\MAGT{^\alpha_\mu_\nu}$ and~$\MAGQ{_{\alpha\mu\nu}}$ leads to the \emph{metric-affine} formulation~\cite{BeltranJimenez:2016wxw,Percacci:2020ddy,Marzo:2021iok,JimenezCano:2021rlu,Delhom:2021bvq}, while the a priori restriction~$\MAGQ{_{\alpha\mu\nu}}\equiv 0$ yields \emph{Einstein-Cartan} gravity~\cite{Cartan:1922,Cartan:1923,Cartan:1924,Cartan:1925,Einstein:1925, Einstein:1928,Einstein:19282}. Excluding both~$\MAGT{^\alpha_\mu_\nu}\equiv\MAGQ{_{\alpha\mu\nu}}\equiv 0$ results in the most-commonly used \emph{metric} formulation of GR~\cite{Einstein:1915}, in which the connection loses its independence from the metric and is fixed by the Christoffel formula~$\MAGA{^\mu_\nu_\rho}\equiv\Con{^\mu_{\nu}_{\rho}}\equiv\tensor{g}{^{\mu\lambda}}\big(\PD{_{(\nu}}\g{_{\rho)\lambda}}-\frac{1}{2}\PD{_{\lambda}}\g{_{\nu\rho}}\big)$. There are many more options~\cite{Einstein:1915,Weyl:1918,Palatini:1919,Weyl:1922,Cartan:1922,Eddington:1923,Cartan:1923,Eddington:1923,Cartan:1924,Cartan:1925,Einstein:1925,Einstein:1928,Einstein:19282,Schroedinger:1950,Moller:1961,Pellegrini:1963,Hayashi:1967se,Cho:1975dh,Hehl:1976kt,Hehl:1976kv,Hehl:1976my,Hehl:1977fj,Kijowski:1978,Hayashi:1979qx,Nester:1998mp,BeltranJimenez:2019odq}, and even interpolations~\cite{Aringazin:1991,Heinicke:2005bp,Baekler:2006vw,BeltranJimenez:2015pnp,BeltranJimenez:2016wxw,BeltranJimenez:2016gnn,BeltranJimenez:2016wuf,BeltranJimenez:2017vop,Iosifidis:2018jwu,Iosifidis:2018diy,Iosifidis:2018zjj,Janssen:2019doc,Janssen:2019doc,Harko:2021tav,Quiros:2023owm} between them: full metric affine geometry is completely unrestricted, and will be our focus.
 \begin{figure}
 	\includegraphics[width=1\linewidth]{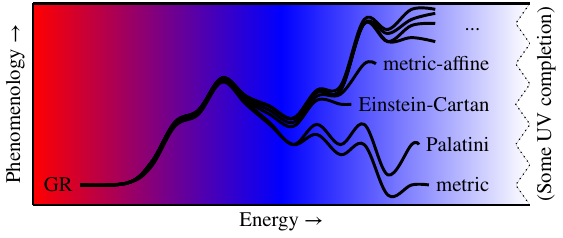}
 	\caption{As a low-energy effective theory, Einstein's GR has many equivalent geometrical formulations that will only be distinguishable by experimental bounds on their high-energy phenomenology, in particular observational constraints from the very early Universe.}
 	\label{TheoryPhoresis}
 \end{figure}
For the simplest action only consisting of the Ricci scalar~$\MAGF{}\equiv\MAGF{^{\mu\nu}_{\mu\nu}}$, the different formulations of GR are fully equivalent in pure gravity as~$\MAGT{^\alpha_\mu_\nu}\approx\MAGQ{_{\alpha\mu\nu}}\approx 0$ dynamically on the shell. As illustrated in~\cref{TheoryPhoresis}, when SM matter fields are included, formulations start to deviate at high energies, \eg due to effective four-fermion vertices~\cite{Kibble:1961ba, Rodichev:1961,Freidel:2005sn,Alexandrov:2008iy,Shaposhnikov:2020frq,Karananas:2021zkl,Rigouzzo:2023sbb} -- see~\cite{Freidel:2005sn,Bauer:2008zj,Poplawski:2011xf,Diakonov:2011fs,Khriplovich:2012xg,Magueijo:2012ug,Khriplovich:2013tqa,Markkanen:2017tun,Carrilho:2018ffi,Enckell:2018hmo,Rasanen:2018fom,BeltranJimenez:2019hrm,Rubio:2019ypq,Shaposhnikov:2020geh,Karananas:2020qkp,Langvik:2020nrs,Shaposhnikov:2020gts,Mikura:2020qhc,Shaposhnikov:2020aen,Kubota:2020ehu,Enckell:2020lvn,Iosifidis:2021iuw,Racioppi:2021ynx,Cheong:2021kyc,Dioguardi:2021fmr,Piani:2022gon,Dux:2022kuk,Rigouzzo:2022yan,Pradisi:2022nmh,Salvio:2022suk,Rasanen:2022ijc,Gialamas:2022xtt,Gialamas:2023emn,Gialamas:2023flv,Piani:2023aof,Poisson:2023tja,Rigouzzo:2023sbb,Barker:2023fem,Karananas:2023zgg,Martini:2023apm,He:2024wqv} for manifold phenomenological implications.

\paragraph*{Consistency of new gravitational particles}  Before leveraging this high-energy phenomenology, however, we have to worry about more basic self-consistency problems which plague the standard next-to-minimal extensions of GR, i.e. extensions to operators \emph{quadratic} in curvature~\cite{Stelle:1977ry,Neville:1978bk,Neville:1979rb,Sezgin:1979zf,Hayashi:1979wj,Hayashi:1980qp,Hecht:1990wn,Hecht:1991jh,Chen:1998ad}. As anticipated from~\cref{FDef}, these operators carry dynamics for~$\MAGA{^\mu_\nu_\rho}$, i.e.~$\MAGT{^\alpha_\mu_\nu}$ and~$\MAGQ{_{\alpha\mu\nu}}$ in~\cref{tqdef} start to propagate. To see the danger here, we first recall that the force of GR is mediated by a spin-two state, and the forces of the SM are all mediated by spin-one states of vectors. But the torsion, Weyl~\cite{Weyl:1921,BeltranJimenez:2014iie,Iosifidis:2018zwo,Iosifidis:2018diy,Helpin:2019vrv,OrejuelaGarcia:2020viw,Ghilencea:2020piz,BeltranJimenez:2020sih,Xu:2020yeg,Yang:2021fjy,Quiros:2021eju,Quiros:2022uns,Yang:2022icz,Burikham:2023bil,Haghani:2023nrm} and co-Weyl~\cite{Iosifidis:2018jwu} vectors 
\begin{equation}\label{AllVectors}
	\MAGT{_\mu}\equiv\MAGT{^\alpha_\mu_\alpha}\;,\quad
	\MAGQ{_\mu}\equiv\MAGQ{_{\mu\alpha}^\alpha}\;,\quad
	\MAGQt{_\mu}\equiv\MAGQ{^\alpha_\mu_\alpha}\;,
\end{equation}
and the pseudo-vector~$\MAGTt{_\mu}\equiv\epsilon{_{\mu\nu\sigma\lambda}}\MAGT{^{\nu\sigma\lambda}}$ account for merely~16 out of the~64 kinematic d.o.f contained in~$\MAGA{^\mu_\nu_\rho}$. All the remaining d.o.f are carried by traceless rank-three tensors~$\MAGt{^\alpha_\mu_\nu}$ and~$\MAGq{^\alpha_\mu_\nu}$ from torsion and non-metricity, respectively. These irreducible parts contain \emph{two spin-two states and one spin-three state}, all of odd parity~\cite{Percacci:2020ddy}. As is well known, physical consistency imposes severe constraints on theories with multiple massless spin-two fields~\cite{Weinberg:1980kq}, and moreover on all massive~\cite{Chang:1966zza,Singh:1974qz} and massless~\cite{Fronsdal:1978rb,Baldazzi:2021kaf} higher-spin theories with non-linear interactions. Even linearly, there are always ghosts and tachyons~\cite{Stelle:1977ry,Neville:1978bk,Neville:1979rb,Sezgin:1979zf,Hayashi:1979wj,Hayashi:1980qp}. It seems unlikely that generic metric-affine models will be consistent. Indeed, extensive research into this question has shown that whilst unwanted particles can be eliminated from the linearised gravity spectrum by tuning~\cite{Sezgin:1981xs,Blagojevic:1983zz,Blagojevic:1986dm,Kuhfuss:1986rb,Yo:1999ex,Yo:2001sy,Blagojevic:2002,Puetzfeld:2004yg,Yo:2006qs,Shie:2008ms,Nair:2008yh,Nikiforova:2009qr,Chen:2009at,Ni:2009fg,Baekler:2010fr,Ho:2011qn,Ho:2011xf,Ong:2013qja,Puetzfeld:2014sja,Karananas:2014pxa,Ni:2015poa,Ho:2015ulu,Karananas:2016ltn,Obukhov:2017pxa,Blagojevic:2017ssv,Blagojevic:2018dpz,Tseng:2018feo,Lin:2018awc,BeltranJimenez:2019acz,Zhang:2019mhd,Aoki:2019rvi,Zhang:2019xek,Jimenez:2019qjc,Lin:2019ugq,Percacci:2019hxn,Barker:2020gcp,BeltranJimenez:2020sqf,MaldonadoTorralba:2020mbh,Barker:2021oez,Marzo:2021esg,Marzo:2021iok,delaCruzDombriz:2021nrg,Baldazzi:2021kaf,Annala:2022gtl}, problems almost inevitably reappear in the full theory due to strongly coupled modes~\cite{Moller:1961,Pellegrini:1963,Hayashi:1967se,Cho:1975dh, Hayashi:1979qx,Hayashi:1979qx,Dimakis:1989az,Dimakis:1989ba,Lemke:1990su,Hecht:1990wn,Hecht:1991jh,Yo:2001sy,Afshordi:2006ad,Magueijo:2008sx,Charmousis:2008ce,Charmousis:2009tc,Papazoglou:2009fj,Baumann:2011dt,Baumann:2011dt,DAmico:2011eto,Gumrukcuoglu:2012aa,Wang:2017brl,Mazuet:2017rgq,BeltranJimenez:2020lee,JimenezCano:2021rlu,Barker:2022kdk,Delhom:2022vae,Annala:2022gtl,Barker:2022kdk,Barker:2023fem} (see strong coupling in other theories~\cite{Vainshtein:1972sx,Deffayet:2001uk,Deffayet:2005ys,Charmousis:2008ce,Charmousis:2009tc,Papazoglou:2009fj,deRham:2014zqa,Deser:2014hga,Wang:2017brl}) and the breaking of accidental symmetries which only exist linearly~\cite{Velo:1969txo,Aragone:1971kh,Cheng:1988zg,Hecht:1996np,Chen:1998ad,Yo:1999ex,Yo:2001sy,Blixt:2018znp,Blixt:2019ene,Blixt:2020ekl,Krasnov:2021zen,Bahamonde:2021gfp,Delhom:2022vae} (see also~\cite{Hayashi:1980qp,Blagojevic:1983zz,Blagojevic:1986dm,Yo:2001sy,Blagojevic:2002,Ong:2013qja,Blagojevic:2013dea,Blagojevic:2013taa,Blagojevic:2018dpz,BeltranJimenez:2019hrm,Aoki:2020rae,Barker:2021oez}). To summarise the current state of the field, the metric-affine formulation seems to be in `tension' with the model-building lessons of quantum field theory~\cite{Puetzfeld:2004yg,Yo:2006qs,Shie:2008ms,Chen:2009at,Ni:2009fg,Baekler:2010fr,Ho:2011qn,Ho:2011xf,Puetzfeld:2014sja,Ho:2015ulu,Ni:2015poa,Tseng:2018feo,Zhang:2019mhd,Zhang:2019xek,BeltranJimenez:2019acz,BeltranJimenez:2019esp,MaldonadoTorralba:2020mbh,Barker:2020gcp,Percacci:2020ddy,BeltranJimenez:2020sqf,delaCruzDombriz:2021nrg,Marzo:2021esg,Piva:2021nyj,Marzo:2021iok,Iosifidis:2021xdx,Jimenez-Cano:2022sds,Iosifidis:2023pvz}.

\paragraph*{In this letter} We shall show that symmetries exist which can bring metric-affine gravity into harmony with particle physics. Our main focus will be on a transformation that is generated by a \emph{pair} of independent vectors. We call this \emph{extended projective} (EP) symmetry, because it contains known projective transformations~\cite{Schroedinger:1950,Trautman1973,Sandberg:1975db,Hehl:1976kv,Trautmann1976,Hehl:1978, Hehl:1981,Afonso:2017bxr,Iosifidis:2018jwu,Janssen:2018exh,Gomes:2022vrc}. The vector 
\begin{equation} \label{VEP}
	\MAGV{_\mu} \equiv 2\MAGT{_\mu}+\MAGQ{_\mu}-\MAGQt{_\mu},
\end{equation}
as well as~$\MAGTt{_\mu}$ are the only EP-two invariant vectors (see explicit definition in~\cref{extendedProjectiveTransformation} below). Our initial motivation for EP-transformations is that exactly~$\MAGTt{_\mu}$ and~$\MAGV{_\mu}$ (eight out of the 64 d.o.f) appear in the general GR coupling to SM fermions, as discussed below. Apart from EP symmetry, we show that metric-affine gravity has one further double-vector symmetry. This is defined by the invariants~$\MAGQ{_\mu}$ and~$\MAGTt{_\mu}$, so we dub it \emph{iso-Weyl} (IW) invariance. As we shall show, all these symmetries are non-accidental, \ie they are valid at all orders in the weak-field expansion. Moreover, a general analysis of all operators shows that the symmetries serve a protective role: they are \emph{precisely those needed} to guarantee self-consistent particle content. These could be early hints that EP invariance is as relevant to the metric-affine formulation as diffeomorphism invariance is to the metric formulation (see approaches in~\cite{Fierz:1939ix,Deser:1969wk,Deser:1987uk,Deser:2001wx,Padmanabhan:2004xk,Butcher:2009ta}). We separately discuss the promising phenomenology of the resulting models.

\paragraph*{Non-dynamical symmetries} First, we shall consider an almost trivial situation without curvature-squared terms. In this case, we can obtain arbitrary vector symmetries by tuning the action appropriately. Even for a triple-vector symmetry, we can form an invariant vector~$\MAGVt{_\mu}$ as any linear combination of the four vectors~$\MAGT{_\mu}$,~$\MAGQ{_\mu}$,~$\MAGQt{_\mu}$, and~$\MAGTt{_\mu}$. Then the action 
\begin{equation} \label{ActionThreeVector}
	S = \int\mathrm{d}^4x\sqrt{-g}\left[\frac{\MPl{}^2}{2}\rR{} + b_1 \MPl{}^2 \MAGVt{_\mu} \MAGVt{^\mu}\right],
\end{equation}
exhibits the triple-vector symmetry, where~$\rR{}$ is the Riemannian Ricci scalar (formed from~$\Con{^\mu_\nu_\rho}$). In~\cref{ActionThreeVector}, we can equivalently replace~$\rR{}$ by the metric-affine Ricci scalar~$\MAGF{}$ if we carefully add terms quadratic in~$\MAGT{_\mu}$,~$\MAGQ{_\mu}$,~$\MAGQt{_\mu}$, and~$\MAGTt{_\mu}$ using the post-Riemannian decomposition of~$\MAGF{}$ (see \eg~\cite{Rigouzzo:2022yan}). However,~\cref{ActionThreeVector} does not contain any new particles. Indeed, in the absence of matter we dynamically get~$\MAGVt{_\mu}\approx 0$ on-shell, and hence~\cref{ActionThreeVector} is fully equivalent to GR. We conclude that triple-vector symmetries are too restrictive.

\paragraph*{Projective symmetry} Next, we turn to the best-known (after diffeomorphism invariance) single-vector symmetry, namely projective invariance. This leaves~$\MAGF{}$ invariant and is defined by the transformation (see~\cite{Rigouzzo:2022yan})
\begin{equation} \label{projectiveTransformation}
	\begin{gathered}
		\MAGT{_\mu} \mapsto \MAGT{_\mu} - 3 \ShiftA{_\mu} \;, \quad \MAGQ{_\mu}\mapsto \MAGQ{_\mu} + 8 \ShiftA{_\mu} \;, \\ \MAGQt{_\mu} \mapsto \MAGQt{_\mu} + 2 \ShiftA{_\mu} \;, \quad \MAGTt{_\mu} \mapsto \MAGTt{_\mu} \;,
	\end{gathered}
\end{equation}
where $\ShiftA{^\mu}$ is a local vector. Unfortunately, the general projective-invariant theory formed from~\cref{tqdef,FDef} is awash with parameters --- see~\cref{ProjTheoryGeneral}. Without loss of generality, we arbitrarily set all these free parameters equal to some number~$\alpha$. This results in a one-parameter projective-invariant extension of minimal GR
\begin{align}
	&S = \int\mathrm{d}^4x\sqrt{-g}\Bigg[\frac{\MPl{}^2}{2}\MAGF{}
	-\alpha\bigg(\MAGF{^{\rho\sigma\mu\nu}}\Big[
	2\MAGF{_{(\rho\sigma)\mu\nu}}
	+\MAGF{_{\mu\nu\rho\sigma}}
	\nonumber\\
	&
	+\MAGF{_{\nu\sigma\mu\rho}}
	+2\MAGF{_{(\nu\rho)\mu\sigma}}
	\Big]
	-2\MAGFb{^{(\mu\nu)}}
	\MAGFb{_{(\mu\nu)}}
	+\MAGFa{^{\mu\nu}}\Big[
	\MAGFa{_{\mu\nu}}
	-19\MAGFa{_{\nu\mu}}
	\Big]
	\nonumber\\
	&
	+\MAGFh{^{\mu\nu}}\Big[
	\MAGFh{_{\mu\nu}}
	-\MAGFb{_{\mu\nu}}
	-11\MAGFa{_{\mu\nu}}
	\Big]
	+\MAGFa{^{\mu\nu}}
	\MAGFb{_{\mu\nu}}
	+\MAGF{}^2\bigg)
	\Bigg] \;,
\label{ProjTheory}
\end{align}
where the Ricci tensor is~$\MAGFb{_{\mu\nu}}\equiv\MAGF{^\lambda_{\mu\lambda\nu}}$, the co-Ricci is~$\MAGFa{_{\mu\nu}}\equiv\MAGF{_{\mu}^\lambda_{\lambda\nu}}$ and~$\MAGFh{_{\mu\nu}}\equiv\MAGF{^\lambda_\lambda_{\mu\nu}}\equiv\tensor{\partial}{_{[\mu}}\MAGQ{_{\nu]}}$ is the homothetic curvature~\cite{BeltranJimenez:2016wxw,Iosifidis:2018jwu}. The saturated propagator of~\cref{ProjTheory} features e.g. spin-two and spin-three poles at~$\MPl{}^2/10\alpha$ and~$-\MPl{}^2/14\alpha$ respectively (see~\cref{SpectralAppendix} for other pathologies). One of these poles is obviously a tachyon, and anyway their spins signal inconsistency. It is straightforward to prove (see supplemental material in~\cite{Barker:2024}) that the inconsistencies are generic; not just a product of our arbitrary tuning in~\cref{ProjTheory}. Thus, projective invariance is not restrictive enough, failing to bootstrap metric-affine gravity. Alternative or additional symmetries are needed, and to help identify these we turn to the matter sector.

\begin{figure}
	\includegraphics[width=1\linewidth]{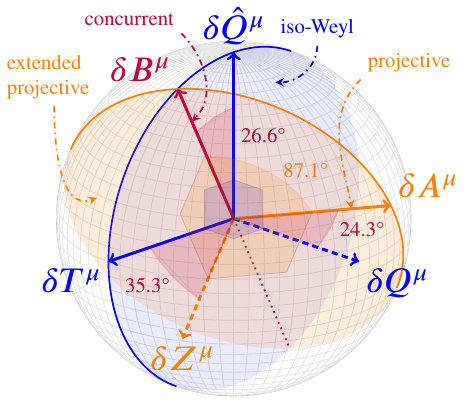}
	\caption{The projective symmetry~\cref{projectiveTransformation} lies in the plane of \emph{extended projective} symmetries, inclined to the torsion-trace, Weyl and co-Weyl axes defined in~\cref{AllVectors}. This plane is normal to the vector~$\MAGV{_\mu}$ in~\cref{VEP}; the only part of metric-affine geometry to couple non-minimally to matter in the Standard Model. The further \emph{iso-Weyl} symmetry in~\cref{isoWeylTransformation} is normal to the Weyl vector~$\MAGQ{_\mu}$, and the \emph{concurrent} symmetry~\cref{symIntersection} lies at their intersection.}
	\label{Fig2}
\end{figure}

\paragraph*{Coupling to matter} Since the kinetic term of scalars does not contain the affine connection and gauge invariance similarly forbids a direct coupling of vector bosons, we shall focus on a fermion~$\psi$ with generic kinetic term~\cite{Freidel:2005sn,Alexandrov:2008iy,Rigouzzo:2023sbb}
\begin{subequations}
\begin{align}
	&\frac{i}{2}\bar{\psi}\big(1-i\alpha-i\beta\gamma^5\big)\gamma^\mu\mathcal{D}_\mu\psi+\text{h.c.}
	\nonumber\\
	&\hspace{30pt}
	\equiv\frac{i}{2}\bar{\psi}\gamma^\mu\mathcal{\mathring{D}}_\mu\psi+\text{h.c.}+\MAGTt{_\mu}\Ja{^\mu}+\MAGV{_\mu}\Jv{^\mu}\label{fermionKinetic},
	\\
	&\Ja{^\mu}\equiv-\frac{1}{8}\bar{\psi}\gamma^5\gamma^\mu\psi,\quad
	\Jv{^\mu}\equiv\frac{\alpha}{4}\bar{\psi}\gamma^\mu\psi-2\beta\Ja{^\mu},\label{CurrentDefinition}
\end{align}
\end{subequations}
where~$\alpha$ and~$\beta$ are two real non-minimal coupling constants, and (M) denotes the minimal (axial) current to which~$\MAGTt{_\mu}$ is well-known to couple, marking it out as a special part of the geometry. The physics of~\cref{fermionKinetic} is our first indication that~$\MAGV{_\mu}$ in~\cref{VEP} captures another very special part of the spacetime geometry: it is the part which couples to the \emph{non}-minimal current (N). For a \emph{minimal} coupling~$\alpha=\beta=0$, we see that~\cref{fermionKinetic} is symmetric under the triple-vector transformation that leaves~$\MAGTt{_\alpha}$ invariant. If~\cref{ActionThreeVector} were an interesting formulation of GR, it could be minimally coupled to the SM by identifying~$\MAGVt{_\mu}=\MAGTt{_\mu}$. However~\cref{ActionThreeVector} is relatively boring, and moreover there is no justification for the~$\alpha=\beta=0$ restriction. Does the non-minimal case motivate any more substantive model? The projective transformation in~\cref{projectiveTransformation} leaves both~$\MAGTt{_\mu}$ and~$\MAGV{_\mu}$ invariant, and indeed it is known that projective invariance is \emph{compatible} with a non-minimal SM coupling in this way. But projective models such as that in~\cref{ProjTheory} are \emph{too} substantive: contrasting starkly with~\cref{ActionThreeVector}, we just showed that the particle spectrum is completely out of control. Yet, just as it inspired the problem,~\cref{fermionKinetic} tells us where to look for the solution: projective invariance is an essentially arbitrary slice of the most general \emph{extended projective} (EP) symmetry that leaves~\cref{fermionKinetic} invariant -- see illustration in~\cref{Fig2}.

\paragraph*{Extended projective symmetry} The transformation 
\begin{equation}\label{extendedProjectiveTransformation}
\begin{gathered}
\MAGT{_\mu}\mapsto\MAGT{_\mu}-3\ShiftA{_\mu}+\ShiftB{_\mu}\;,\quad\MAGQ{_\mu}\mapsto\MAGQ{_\mu}+8\ShiftA{_\mu}\;,\\\quad\MAGQt{_\mu}\mapsto\MAGQt{_\mu}+2\ShiftA{_\mu}+2\ShiftB{_\mu}\;,  \quad \MAGTt{_\mu} \mapsto \MAGTt{_\mu} \;,
\end{gathered}
\end{equation}
corresponding to~$\MAGA{_\mu_\nu_\rho}\mapsto \MAGA{_\mu_\nu_\rho} + \big[\g{_{\mu \rho}}(2 \ShiftB{_\nu} - 5 \ShiftA{_\nu}) + 2 \g{_{\nu (\mu}}(\ShiftA{_{\rho)}} - 4 \ShiftB{_{\rho)}})\big]/36$, defines EP symmetry, where~$\ShiftA{_\mu}$ and~$\ShiftB{_\mu}$ are two arbitrary vector fields. Clearly,~\cref{extendedProjectiveTransformation} is the most general transformation consistent with the SM in~\cref{fermionKinetic} and moreover contains the projective case,~\cref{projectiveTransformation}, with~$\ShiftB{_\mu}=0$. It is also distinct from previously identified metric-affine vector symmetries~\cite{Iosifidis:2019fsh,Percacci:2020ddy} (see also~\cite{Garcia-Parrado:2020lpt}). Amazingly, EP symmetry admits dynamics besides the graviton. The Holst term
\begin{equation} \label{Holst}
	\MAGFt{}\equiv\epsilon^{\mu\nu\sigma\lambda}\MAGF{_{\mu\nu\sigma\lambda}}=\frac{1}{3}\MAGTt{_\mu}\MAGV{^\mu}-\rD{_\mu}\MAGTt{^\mu}+[(t,q)^2]\;,
\end{equation}
is an invariant of EP transformations (see~\cite{Rigouzzo:2023sbb} for the second equality). Here~$[(t,q)]^2$ stands for a specific linear combination of all seven possible terms quadratic in the pure tensor parts (see~\cite{Rigouzzo:2022yan}). The most general EP-invariant theory with quadratic operators is (see supplemental materials~\cite{Barker:2024})
\begin{align}
	S&=\int\mathrm{d}^4x\sqrt{-g}\Big[
	\MPl{}^2 \Big(\frac{1}{2}\rR{}
	+b_1\MAGV{_\mu}\MAGV{^\mu}	+b_2\MAGTt{_\mu}\MAGTt{^\mu}\nonumber\\
	& +b_3\MAGTt{_\mu}\MAGV{^\mu}+[(t,q)^2] \Big)+ \alpha\MAGFt{}^2 + \MAGTt{_\mu}\Ja{^\mu}+\MAGV{_\mu}\Jv{^\mu}\Big] \;.
	\label{ActionExtendedProjective}
\end{align}
When we discuss \emph{general} models we use~\cref{tqdef,FDef} as our only ingredients, because only~$\MAGF{^\rho_{\sigma\mu\nu}}$,~$\MAGT{^\alpha_{\mu\nu}}$ and~$\MAGQ{_{\alpha\mu\nu}}$ have geometric meaning. We reject operators with explicit derivatives: inclusion of these operators maps metric affine gravity to the unresolved problem of finding a consistent nonlinear rank-three field theory~\cite{Chang:1966zza,Singh:1974qz,Fronsdal:1978rb,Baldazzi:2021kaf}. By adding appropriate terms quadratic in~$\MAGT{_\mu}$,~$\MAGTt{_\mu}$,~$\MAGQ{_\mu}$ and~$\MAGQt{_\mu}$, we replaced~$\MAGF{}$, which fails to be EP-invariant, by Riemannian~$\rR{}$. Also, we did not include a linear power of~$\MAGFt{}$, which could be reabsorbed by a redefinition of~$b_3$, up to a total derivative. Starting with~\cite{Hecht:1996np}, models with~$\MAGFt{}^2$ similar to~\cref{ActionExtendedProjective} were studied in~\cite{BeltranJimenez:2019hrm,Pradisi:2022nmh,Salvio:2022suk,Gialamas:2022xtt} (without our symmetry motivations, or metric-affine geometry). Introducing an auxiliary field~$\phi$ and expanding the Holst term as shown in~\cref{Holst},~\cref{ActionExtendedProjective} becomes~\cite{BeltranJimenez:2019hrm,Pradisi:2022nmh,Salvio:2022suk,Gialamas:2022xtt}
\begin{align}
	&S=\int\mathrm{d}^4x\sqrt{-g}\Big[\MPl{}^2\Big(\frac{1}{2}\rR{}
	-\alpha \phi^2  	+b_1\MAGV{_\mu}\MAGV{^\mu} 	+b_2\MAGTt{_\mu}\MAGTt{^\mu}
	\nonumber\\
	&
 +g(\phi)\MAGTt{_\mu}\MAGV{^\mu} \Big)	-2\alpha \MPl{} \phi \rD{_\mu} \MAGTt{^\mu}
+\MAGTt{_\mu}\Ja{^\mu}+\MAGV{_\mu}\Jv{^\mu}\Big] \;,
	\label{ActionExtendedProjectiveAux}
\end{align}
where~$g(\phi)\equiv\CouplingB{3}+2\alpha\phi/3\MPl{}$ and we already left out the higher-spin tensor parts since they only appear quadratically and hence are not sourced, \ie vanish dynamically as~$\MAGt{^\mu_{\nu\sigma}}\approx\MAGq{^\mu_{\nu\sigma}}\approx 0$ on the shell~\cite{Rigouzzo:2022yan,Rigouzzo:2023sbb}. Now we solve for the vectors~$\MAGV{_\mu}$ and~$\MAGTt{_\mu}$ and plug the result back in~\cref{ActionExtendedProjectiveAux} to obtain an equivalent theory in the metric formulation of GR
%Using the formulae of~\cite{BeltranJimenez:2019hrm,Rigouzzo:2023sbb}, we get
\begin{align}
	S=\int\mathrm{d}^4x\sqrt{-g}\bigg[&\frac{\MPl{}^2}{2}\rR{}
	-\frac{f(\phi)}{2}\PD{_\mu}\phi\PD{^\mu}\phi 
	\nonumber\\
	&
	-\alpha \MPl{}^2\phi^2+\mathcal{L}_{\psi}\bigg],
	\label{ActionExtendedProjectiveFinal}
\end{align}
where~$f(\phi)\equiv 8b_1\alpha^2/\big(4\CouplingB{1}\CouplingB{2}-g(\phi)^2\big)$ and~$\mathcal{L}_{\psi}$ contains the matter. The conclusion (for~$b_1\neq 0$) is that EP invariance leads to a new pseudoscalar.

\paragraph*{Inflation \& dark matter} Next, we transform~$\phi$ to a canonically normalized scalar field~$\chi\equiv\chi(\phi)$. The result depends on the signs of~$b_1$ and~$b_2$ (see~\cite{BeltranJimenez:2019hrm}) -- for the important example~$b_1<0$,~$b_2>0$, we get from~\cref{ActionExtendedProjectiveFinal}
\begin{subequations} \label{canonicalAction}
\begin{align} 
	S&=\int\mathrm{d}^4x\sqrt{-g}\Big[\frac{\MPl{}^2}{2}\rR{}-\frac{1}{2} \PD{_\mu}\chi\PD{^\mu}\chi -U+\mathcal{L}_{\psi}\Big],\label{EvidentScalar}\\
	U&\equiv\frac{9 \MPl{}^4}{\alpha} \bigg[\sqrt{|b_1 b_2|} \sinh\bigg(\frac{\chi+\chi_0}{3 \sqrt{2|b_1|} \MPl{}}\bigg)- \frac{b_3}{2}\bigg]^2,\label{Potential}
\end{align}
\end{subequations}
where~$\chi_0\equiv 3 \sqrt{2|b_1|}\MPl{}\,\text{arcsinh}\big(b_3/2\sqrt{|b_1 b_2|}\big)$. It was shown in~\cite{Salvio:2022suk,Gialamas:2022xtt,He:2024wqv}, that potentials similar to~\cref{Potential} can drive inflation. In particular, there is a region of parameter space with large values of~$b_3$,  in which the potential develops a plateau and predictions are very close to the universal values~$n_s\approx 1-2/N$ for the spectral index and~$r\approx12/N^2$ for the tensor-to-scalar ratio~\cite{Salvio:2022suk} (with ~$N\approx 55$ the number of e-foldings needed for cosmic microwave background perturbations). These predictions agree excellently with current cosmological observations~\cite{Planck:2018jri, BICEP:2021xfz}, which will be refined with upcoming precision measurements~\cite{arXiv:1110.3193,Weltman:2018zrl,Abazajian:2019eic,LiteBIRD:2022cnt}. Dedicated analysis of~\cref{Potential} is pending. Secondly, and reverting to~$\phi\equiv\phi(\chi)$, the effective summands in~$\mathcal{L}_{\psi}$ are
\begin{subequations}
\begin{align}
	\scalebox{.4}{\raisebox{-9mm}{\includegraphics{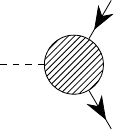}}\hspace{4mm}}
	%\scalebox{.5}{\raisebox{-8mm}{\includegraphics{ThreeVertex.pdf}}\hspace{3mm}}
	%&\scalebox{.5}{$\feynmandiagram[small, inline=(b.base), horizontal=d to b]{a -- [fermion] b [blob] -- [fermion] c,b -- [scalar] d,}$}
	&\hspace{-5pt}=\frac{f(\phi)\Big(g(\phi)\Jv{^\mu}-2\CouplingB{1}\Ja{^\mu}\Big)\PD{_\mu}\phi}{4\alpha b_1\MPl{}},\label{Process1}
	\\
	%&\scalebox{.4}{$\feynmandiagram[small, inline=(b.base), horizontal=a to d]{a -- [fermion] b [blob] -- [fermion] c,e -- [fermion] b -- [fermion] d,}$}
	\scalebox{.4}{\raisebox{-10mm}{\includegraphics{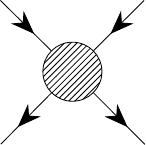}}}
	&\hspace{-5pt}=\frac{f(\phi)\Big[\left(g(\phi)\Jv{_\mu}-b_1\Ja{_\mu}\right)\Ja{^\mu}-b_2\Jv{_\mu}\Jv{^\mu}\Big]}{8\alpha^2b_1\MPl{}^2},\label{Process2}
\end{align}
\end{subequations}
where~\cref{Process1} is a scalar-fermion vertex and~\cref{Process2} represents a four-Fermi interaction (see~\cite{Shaposhnikov:2020frq,Karananas:2021zkl,Rigouzzo:2023sbb}). Both vertices can lead to the production of fermionic dark matter~\cite{Shaposhnikov:2020aen, Rigouzzo:2023sbb}.

\paragraph*{Einstein-Cartan link} As a final comment on EP invariance, note how~\cref{extendedProjectiveTransformation} can set~$\MAGQ{^\mu},\MAGQt{^\mu}\mapsto 0$. Since~$\MAGq{^\mu_{\nu\sigma}}\approx 0$ dynamically, the whole of~$\MAGQ{_{\lambda\mu\nu}}$ is then removed. This pure-torsion gauge, in which~$\MAGV{_\mu}\mapsto 2\MAGT{_\mu}$ in~\cref{VEP}, effectively reduces the metric-affine formulation to Einstein-Cartan gravity (where~$\MAGQ{_{\alpha\mu\nu}}\equiv 0$ a priori). We find this observation extremely interesting since, as is well-known, it is possible to derive Einstein-Cartan gravity from gauging spinorial Poincar\'e invariance~\cite{Utiyama:1956sy,Kibble:1961ba,Sciama:1962}. Our findings thus establish a second link between spinor fields and Einstein-Cartan geometry. We see that fermions play a special role in gravity and hint towards curvature and torsion, but no non-metricity.

\paragraph*{An alternative double-vector symmetry} Before concluding, we point out that there is one more double-vector symmetry. We define \emph{iso-Weyl} (IW) invariance
	\begin{equation} \label{isoWeylTransformation}
		\MAGT{_\mu}\mapsto\MAGT{_\mu}+\ShiftB{_\mu}\;,\quad\MAGQt{_\mu}\mapsto\MAGQt{_\mu}+2\ShiftB{_\mu}+\ShiftC{_\mu}\;,
\end{equation}
with inert~$\MAGQ{_\mu}$ and~$\MAGTt{_\mu}$ and local~$\ShiftC{^\mu}$. In~\cref{app:theories} we show that the most general IW-invariant action has a Maxwell-like term in homothetic curvature~$\MAGFh{_{\mu\nu}}\equiv\tensor{\partial}{_{[\mu}}\MAGQ{_{\nu]}}$, leading to Einstein--Proca theory~\cite{BeltranJimenez:2014iie,Iosifidis:2018diy,Iosifidis:2018zwo,Helpin:2019vrv,OrejuelaGarcia:2020viw,Ghilencea:2020piz,BeltranJimenez:2020sih,Xu:2020yeg,Yang:2021fjy,Quiros:2021eju,Quiros:2022uns,Yang:2022icz,Burikham:2023bil,Haghani:2023nrm} (see also~\cite{Aringazin:1991,Vitagliano:2010pq,Iosifidis:2021xdx}). We can make sense of the particle ramifications of EP and IW symmetries in the following way. A vector~$\tensor{V}{_\mu}$ admits only~$\PD{_{[\mu}}\tensor{V}{_{\nu]}}\PD{^{[\mu}}\tensor{V}{^{\nu]}}$ and~$(\PD{_{\mu}}\tensor{V}{^{\mu}})^2$ as viable kinetic operators; the latter is often overlooked, but propagates a healtly scalar if a mass is present. Now in hindsight, it is obvious that~$\MAGFh{_{\mu\nu}}\MAGFh{^{\mu\nu}}$ and~$\MAGFt{}^2$ contain these operators. What is \emph{not} obvious, is that these operators essentially follow from IW and EP symmetries, or that~$\MAGt{^\mu_{\nu\sigma}}$ and~$\MAGq{^\mu_{\nu\sigma}}$ drop out of the resulting models. We conjecture that \emph{no further vector symmetries will be found which give rise to consistent new particles in metric-affine gravity}.

\paragraph*{Concluding remarks} The metric-affine formulation of General Relativity stands out since it does not require any a priori assumptions about the vanishing of curvature, torsion or non-metricity. However, it has a crucial problem: next-to-minimal models that include curvature-squared terms are generically plagued by inconsistencies, among others caused by higher-spin particles. We have addressed these issues with new symmetries. Going beyond the commonly-considered single-vector transformations, we have shown that \emph{extended projective} (EP) symmetry, a double-vector invariance generalizing known projective transformations, achieves three goals;
\begin{enumerate}
	\item General metric-affine gravity contains numerous a priori undetermined parameters. This leads to a significant ambiguity and loss of predictive power. Imposing EP symmetry leaves us with very few parameters.
	\item We have constructed the most general EP-invariant theory that is composed of terms that are at most quadratic in curvature, torsion and non-metricity. As our key outcome, we have demonstrated that this model is free of the notorious inconsistencies.
	\item  The general EP model addresses open phenomenological problems with great predictive economy. Firstly, it introduces exactly one new heavy scalar, with potential \eqref{Potential} suitable for inflation. Secondly, the processes shown in~\cref{Process1,Process2} enable a well-established mechanism for fermionic dark matter production.
\end{enumerate} 

Crucially, EP invariance was not \emph{engineered for} the specific observables in~\cref{Potential,Process1,Process2}, but instead \emph{derived from} the matter coupling. Namely it is the full symmetry of a non-minimal Dirac sector. We note an alternative \emph{iso-Weyl} (IW) double-vector symmetry: this propagates a heavy vector, for which there is little phenomenological demand. In summary therefore, EP rather than IW invariance is indicated by the Occam principle, and we put forwards EP-invariant GR as the natural conclusion of metric-affine gravity.

\begin{acknowledgments}

We are grateful for useful discussions with Will Handley, Mike Hobson, Damianos Iosifidis, Georgios Karananas, Anthony Lasenby, Carlo Marzo, Roberto Percacci and Claire Rigouzzo.

	W.~B. is grateful for the kind hospitality of Leiden University and the Lorentz Institute, and the support of Girton College, Cambridge. S.~Z.~acknowledges support of the \emph{Fonds de la Recherche Scientifique} -- FNRS.

This work used the Newton server, access to which was provisioned by Will Handley, and funded through an ERC grant.

This work used the DiRAC Data Intensive service (CSD3 \href{www.csd3.cam.ac.uk}{www.csd3.cam.ac.uk}) at the University of Cambridge, managed by the University of Cambridge University Information Services on behalf of the STFC DiRAC HPC Facility (\href{www.dirac.ac.uk}{www.dirac.ac.uk}). The DiRAC component of CSD3 at Cambridge was funded by BEIS, UKRI and STFC capital funding and STFC operations grants. DiRAC is part of the UKRI Digital Research Infrastructure.
\end{acknowledgments}

\appendix

\section{Conventions} \label{app:conventions} 
In this appendix we summarize our conventions. We work in natural units~$c\equiv\hbar\equiv 1$ with the reduced Planck mass~$\MPl{}=1/\sqrt{8 \pi G}$, where~$G$ is Newton's constant. Moreover, we use the Minkowski metric~$\tensor{\eta}{_{\mu\nu}}=(-,+,+,+)$ and the covariant derivative of a vector~$\tensor{V}{^\nu}$ is defined as~$\rcD{_\mu} V^\nu =\PD{_\mu} \tensor{V}{^\nu} + \MAGA{^\nu_\mu_\rho} \tensor{V}{^\rho}$. Relevant contractions of curvature~\cref{FDef} are the Ricci scalar~$\MAGF{}\equiv\MAGF{^{\mu\nu}_{\mu\nu}}$ and the homothetic curvature~$\MAGFt{_{\mu\nu}}\equiv\MAGF{^\alpha_{\alpha\mu\nu}}$. For the gamma matrices, we use the conventions~$\left\{\gamma_A, \gamma_B \right\} = - 2 \eta_{AB}$ and~$\gamma_5 = -i \gamma^0 \gamma^1 \gamma^2 \gamma^3 = i \gamma_0 \gamma_1 \gamma_2 \gamma_3$, where the Levi-Civita tensor is~$\epsilon_{0123}=1=-\epsilon^{0123}$. For a tensor~$\tensor{T}{_{\mu\nu}}$, square brackets indicate antisymmetrization,~$\tensor{T}{_{[\mu\nu]}}\equiv\frac{1}{2}\big(\tensor{T}{_{\mu\nu}}-\tensor{T}{_{\nu\mu}}\big)$, while round brackets denote symmetrization,~$\tensor{T}{_{(\mu\nu)}}\equiv\frac{1}{2}\big(\tensor{T}{_{\mu\nu}}+\tensor{T}{_{\nu\mu}}\big)$. 

\section{Detailed tree-level particle spectra}\label{SpectralAppendix}

In this appendix we quote the spectra of quantum particles associated with the theories in~\cref{ActionThreeVector,ProjTheory,ActionExtendedProjective,ActionHomothetic,ActionConcurrent}. The analysis is done by brute force using the full set of spin-parity projection operators associated with the 74 d.o.f in metric-affine geometry. The saturated propagator is obtained by computing all gauge symmetries in each case. The positions and residues of propagator poles associated with massive states are extracted, and the massless states obtained in the limit of an explicit lightcone coordinate basis. The analysis required~$\sim 10^3$ core-hours divided between one dedicated compute node consisting of 112 Intel\textsuperscript{\textregistered} \emph{Sapphire Rapids} CPUs supplemented with 64 AMD\textsuperscript{\textregistered} \emph{Ryzen Threadripper} CPUs. The software used for these calculations was applied previously in~\cite{Barker:2024ydb,Barker:2023bmr}.

The general projective-invariant theory is given by~\cite{Percacci:2020ddy}
\begin{align}
	S & = \int\mathrm{d}^4x\sqrt{-g}\Bigg[\frac{\MPl{}^2}{2}\MAGF{}
	-\MAGF{^{\rho\sigma\mu\nu}}\Big(
	\Coupling{1}\MAGF{_{\rho\sigma\mu\nu}}
	+\Coupling{2}\MAGF{_{\sigma\rho\mu\nu}}
	\nonumber\\
	&
	+\Coupling{3}\MAGF{_{\mu\nu\rho\sigma}}
	+\Coupling{4}\MAGF{_{\nu\sigma\mu\rho}}
	+\Coupling{5}\MAGF{_{\nu\rho\mu\sigma}}
	+\Coupling{6}\MAGF{_{\rho\nu\mu\sigma}}
	\Big)
	\nonumber\\
	&
	-\MAGFb{^{\mu\nu}}\Big(
	\Coupling{7}\MAGFb{_{\mu\nu}}
	+\Coupling{8}\MAGFb{_{\nu\mu}}
	\Big)	
	-\MAGFa{^{\mu\nu}}\Big(
	\Coupling{9}\MAGFa{_{\mu\nu}}
	\nonumber\\
	&
	-\big(4\Coupling{1}+4\Coupling{2}-\Coupling{4}+\Coupling{6}+\Coupling{7}-\Coupling{8}-\Coupling{9}+16\Coupling{13}\big)\MAGFa{_{\nu\mu}}
	\Big)
	\nonumber\\
	&
	-\MAGFa{^{\mu\nu}}\Big(
	\Coupling{11}\MAGFb{_{\mu\nu}}
	+\big(\Coupling{5}+2(\Coupling{6}+\Coupling{7}-\Coupling{8})+\Coupling{11}-4\Coupling{14}\big)\MAGFb{_{\nu\mu}}
	\Big)
	\nonumber\\
	&
	-\MAGFh{^{\mu\nu}}\Big(
	\Coupling{13}\MAGFh{_{\mu\nu}}
	-\Coupling{14}\MAGFb{_{\mu\nu}}
	-\big(2\Coupling{1}+2\Coupling{2}+8\Coupling{13}-\Coupling{14}\big)\MAGFa{_{\mu\nu}}
	\Big)
	\nonumber\\
	&
	-\Coupling{16}\MAGF{}^2
	-\MAGT{^{\rho\mu\nu}}\Big(
	\CouplingB{1}\MAGT{_{\rho\mu\nu}}
	+\CouplingB{2}\MAGT{_{\nu\mu\rho}}
	\Big)
	-\CouplingB{3}\MAGT{^{\mu}}\MAGT{_{\mu}}
	\nonumber\\
	&
	-\MAGQ{^{\rho\mu\nu}}\Big(
	\CouplingB{4}\MAGQ{_{\rho\mu\nu}}
	+\CouplingB{5}\MAGQ{_{\nu\mu\rho}}
	\Big)
	-\CouplingB{6}\MAGQ{^{\mu}}\MAGQ{_{\mu}}
	-\CouplingB{7}\MAGQt{^{\mu}}\MAGQt{_{\mu}}
	\nonumber\\
	&
	-\bigg(
	\frac{3\CouplingB{1}}{8}
	+\frac{3\CouplingB{2}}{16}
	+\frac{9\CouplingB{3}}{16}
	-\CouplingB{4}
	-\frac{\CouplingB{5}}{4}
	-4\CouplingB{6}
	-\frac{\CouplingB{7}}{4}
	\bigg)\MAGQ{^{\mu}}\MAGQt{_{\mu}}
	\nonumber\\
	&
	-\CouplingB{9}\MAGQ{_{\mu\rho\nu}}\MAGT{^{\mu\rho\nu}}
	-\bigg(
	\frac{\CouplingB{1}}{4}
	+\frac{\CouplingB{2}}{8}
	+\frac{3\CouplingB{3}}{8}
	+\frac{2\CouplingB{4}}{3}
	-\frac{2\CouplingB{5}}{6}
	+\frac{8\CouplingB{6}}{3}
	\nonumber\\
	&
	-\frac{\CouplingB{7}}{6}
	\bigg)\MAGT{^\mu}\MAGQ{_\mu}
	-\bigg(
		\CouplingB{1}
	+\frac{\CouplingB{2}}{2}
	+\frac{2\CouplingB{3}}{2}
	-\frac{8\CouplingB{4}}{3}
	-\frac{2\CouplingB{5}}{3}
	-\frac{32\CouplingB{6}}{3}
	\nonumber\\
	&
	+\frac{2\CouplingB{7}}{3}
	-\frac{\CouplingB{9}}{3}
		\bigg)\MAGT{^\mu}\MAGQt{_\mu}	
	\Bigg].
\label{ProjTheoryGeneral}
\end{align}
But~\cref{ProjTheoryGeneral} has so many parameters that the formulae for the masses and, in particular, the pole residues, are cumbersome even to write down (see supplemental materials at~\cite{Barker:2024}). To restrict~\cref{ProjTheoryGeneral} to the representative theory in~\cref{ProjTheory} we focus only on the curvature operators by setting~$\CouplingB{1}=\CouplingB{2}=\dots =\CouplingB{7}=\CouplingB{9}=0$. Then, we introduce the dimensionless coupling~$\alpha$ and set~$\Coupling{1}=\Coupling{2}=\dots =\Coupling{9}=\Coupling{11}=\Coupling{13}=\Coupling{15}=\Coupling{16}=\alpha$. The particle spectrum of~\cref{ProjTheory} is given in~\cref{SpectrumProjective}.

\begin{table}
	\caption{\label{SpectrumProjective} The spectrum of massive particles with spin-parity~$J^P$ in the theory in~\cref{ProjTheory}, including the graviton. Note the various inconsistencies in the model.}
\begin{center}
\begin{tabularx}{\linewidth}{l|l|l}
\hline\hline
$J^P$ & Square mass & No-ghost condition from pole residue \\
\hline
$0^-$ &~$\MPl{}^2/2\alpha>0$ &~$\alpha<0$ \\
$1^+$ &~$\MPl{}^2/58\alpha>0$ &~$\alpha<0$ \\
$1^-$ &~$\MPl{}^2/66\alpha>0$ &~$\alpha>0$ \\
$2^-$ &~$\MPl{}^2/10\alpha>0$ &~$\alpha>0$ \\
$3^-$ &~$-\MPl{}^2/14\alpha>0$ &~$\alpha>0$ \\
\hline\hline
\end{tabularx}
\end{center}
\end{table}

In addition to the projective-invariant theory, we also computed the spectra of the theory in~\cref{ActionExtendedProjective} and the most general theory associated with IW invariance as defined in~\cref{isoWeylTransformation} --- these spectra are provided in~\cite{Barker:2024}.

\section{Details about iso-Weyl and concurrent invariance} \label{app:theories}
The general IW-invariant action from~\cref{isoWeylTransformation} is
\begin{align}
	S=\int\mathrm{d}^4x&\sqrt{-g}\Big[
\MPl{}^2\Big(\frac{1}{2}\rR{} + b_1\MAGQ{_\mu}\MAGQ{^\mu}+b_2\MAGTt{_\mu}\MAGTt{^\mu}\nonumber\\
&+b_3\MAGTt{_\mu}\MAGQ{^\mu}+[(t,q)^2]\Big) +\alpha\MAGFh{_{\mu\nu}}\MAGFh{^{\mu\nu}}\Big]
\;,\label{ActionHomothetic}
\end{align}
where~$\MAGFh{_{\mu\nu}}\equiv\tensor{\partial}{_{[\mu}}\MAGQ{_{\nu]}}$ is the homothetic curvature~\cite{BeltranJimenez:2016wxw,Iosifidis:2018jwu}. It is evident from the outset that~$\MAGFh{_{\mu\nu}}$ generates the kinetic term for the vector~$\MAGQ{_\mu}$ and that pure tensor parts are not sourced. Solving for the non-dynamical field yields~$\MAGTt{_\mu} \approx - b_3 \MAGQ{_\mu}/(2 b_2)$, so 
\begin{align}
S=\int\mathrm{d}^4x\sqrt{-g}\Bigg[&
\frac{\MPl{}^2}{2}\rR{}
+\alpha\MAGFh{_{\mu\nu}}\MAGFh{^{\mu\nu}}\nonumber\\
&+ \MPl{}^2\left(b_1-\frac{b_3^2}{4b_2}\right)\MAGQ{_\mu}\MAGQ{^\mu}\Bigg].
\label{ActionHomotheticFinal}
\end{align}
This is Einstein-Proca, or Einstein-Maxwell if~$4b_1b_2=b_3^2$.

Finally, we turn to the intersection of EP and IW invariance, which we shall dub \emph{concurrent} symmetry. It is defined by the transformation
\begin{equation} \label{symIntersection}
	\MAGT{_\mu}\mapsto\MAGT{_\mu}+\ShiftB{_\mu}\;,\quad\MAGQt{_\mu}\mapsto\MAGQt{_\mu}+2\ShiftB{_\mu}\;,
\end{equation}
as follows from \eqs \eqref{extendedProjectiveTransformation} and~\eqref{isoWeylTransformation}. Since this is only a single-vector transformation, the class of models that obey it is larger:
\begin{align}
	S=\int\mathrm{d}^4x\sqrt{-g}\Big[&\MPl{}^2\Big(\frac{1}{2}\rR{}
	+b_1\MAGQ{_\mu}\MAGQ{^\mu} +b_2\MAGTt{_\mu}\MAGTt{^\mu}
	+b_3\MAGQ{_\mu}\MAGTt{^\mu}\nonumber\\
	&+b_4\MAGV{_\mu}\MAGV{^\mu} + b_5\MAGV{_\mu}\MAGQ{^\mu}+b_6\MAGV{_\mu}\MAGTt{^\mu}\nonumber\\
	&+[(t,q)^2]\Big) 	+\alpha\MAGFh{}^2+\tilde{\alpha}\MAGFh{_{\mu\nu}}\MAGFh{^{\mu\nu}}\Big],
	\label{ActionConcurrent}
\end{align}
where we left out the coupling to matter and now the special invariant~$\MAGV{^\mu}$ is re-defined as~$\MAGV{_\mu}=2\MAGT{_\mu}-\MAGQt{_\mu}$. In full analogy to our previous analysis, we introduce the auxiliary field~$\phi$, perform a post-Riemannian decomposition, eliminate the pure tensor parts and non-propagating vectors~$\MAGV{_\mu}$ and~$\MAGTt{_\mu}$, yielding 
\begin{align} \label{ActionConcurrentFinal}
S=\int\mathrm{d}^4x\sqrt{-g}\Big[&\frac{\MPl{}^2}{2}\rR{}
- \alpha \MPl{}^2 \phi^2 + \tilde{\alpha} \MAGFt{_{\mu\nu}} \MAGFt{^{\mu\nu}} \nonumber\\
&-\frac{1}{2} K(\phi) \partial_\mu \phi \partial^\mu \phi+m_Q^2(\phi) \MAGQ{_\mu}\MAGQ{^\mu}\Big],
\end{align}
where  we performed a shift of~$\MAGQ{_\mu}$ proportional to~$\partial_\mu \phi$ in order to remove a mixing term~$\MAGQ{_\mu} \partial_\mu \phi$ and defined
\begin{widetext}
	\begin{subequations}
	\begin{align}
		K(\phi) & \equiv  \frac{2 b_4  \alpha^2}{b_4 b_2 -  \left(\frac{b_6}{2} + \frac{\alpha\phi}{3\MPl{}}\right)^2}  
		-\frac{18 \alpha^2 M_P^4 \left(b_4 \left(4 \alpha  \phi +6 b_3 M_P\right)-b_5 \left(2
			\alpha  \phi +3 b_6 M_P\right)\right){}^2}{	m_Q^2(\phi) \left(4 \alpha ^2 \phi ^2+12 \alpha  b_6
			\phi  M_P+9 \left(b_6^2-4 b_2 b_4\right) M_P^2\right){}^2} \;, \\
%		
%		
%		-\Big[9 \alpha ^2 \MPl{}^4 \left(b_4 \left(4 \alpha  \phi +6 b_3 \MPl{}\right)+b_5 \left(2 \alpha  \phi +3 b_6
%		\MPl{}\right)\right){}^2\Big]
%		\nonumber\\&\ \ \ \ 
%		\times \bigg[2 \big(\MPl{}^2 \big(b_5 \big(4 \alpha ^2 \phi ^2+6 \alpha  \big(b_3
%		+b_6\big) \phi \MPl{}
%		+9 \left(b_2 b_5+b_3 b_6\right) \MPl{}^2\big)
%		+b_4 \left(2 \alpha  \phi +3 b_3 \MPl{}\right){}^2\big)
%		\nonumber\\&\ \ \ \ 
%		+b_1
%		\left(4 \alpha ^2 \phi ^2+12 \alpha  b_6 \phi  \MPl{}+9 \left(b_6^2-4 b_2 b_4\right) \MPl{}^2\right)\big){}^2\bigg]^{-1}, \\
		m_Q^2(\phi) & \equiv  \MPl{}^2\Bigg[b_1 + \Big(\big(b_5 \big(4 \alpha ^2 \phi ^2+6 \alpha  \left(b_3+b_6\right) \frac{\phi}{\MPl{}}
		+9 \left(b_3 b_6 - b_2
		b_5\right) \big)-b_4 \left(2 \alpha  \frac{\phi}{\MPl{}} +3 b_3\right)^2\big)\Big)
		\nonumber\\&\ \ \ \ 
		\times \left(36\left(b_4 b_2 -  \left(\frac{1}{2} b_6  + \frac{1}{3} \alpha \frac{\phi}{\MPl{}}\right)^2\right)\right)^{-1} \Bigg] \;.
	\end{align}
	\end{subequations}
\end{widetext}
	We see that for generic parameter choices both the scalar field~$\phi$ and the vector~$\MAGQ{_\mu}$ are dynamical. The scalar is massive and in general the vector is, too, but for special choices we can have~$m_Q=0$. Both fields can be made healthy for appropriate parameter choices.

\bibliography{NotINSPIRE,Manuscript}

\end{document}